\def\gtorder{\mathrel{\raise.3ex\hbox{$>$}\mkern-14mu
             \lower0.6ex\hbox{$\sim$}}}
\def\ltorder{\mathrel{\raise.3ex\hbox{$<$}\mkern-14mu
             \lower0.6ex\hbox{$\sim$}}}

\documentstyle[12pt,aaspp4]{article}
 \begin{document}

\title{{\it Hubble Space Telescope}
Ultraviolet Images of Five Circumnuclear Star-Forming Rings \footnote
{Based on observations with the {\it Hubble Space Telescope}, which is operated
by AURA, Inc. under NASA contract NAS 5-26555}
}

\author{Dan Maoz}
\affil{School of Physics \& Astronomy and Wise Observatory,
Tel-Aviv University,\\
 Tel-Aviv 69978, Israel,
dani@wise.tau.ac.il}
\author{Aaron J. Barth}
\affil{Department of Astronomy, University of California, Berkeley, CA 94720-3411}
\author{Amiel Sternberg}
\affil{School of Physics \& Astronomy,
Tel-Aviv University, Tel-Aviv 69978, Israel}
\author{Alexei V. Filippenko}
\affil{Department of Astronomy, University of California, Berkeley, CA 94720-3411}
\author{Luis C. Ho}
\affil{Department of Astronomy, University of California, Berkeley, CA 94720-3411,\\
and  Center for Astrophysics, 60 Garden Street, Cambridge, MA 02138}
\author{F.~Duccio Macchetto}
\affil{Space Telescope Science Institute, 3700 San Martin Dr., Baltimore, MD 21218}
\author{Hans-Walter Rix}
\affil{Max-Planck-Institut f\"ur Astrophysik, Karl-Schwarzschild-Strasse 1, D-85740
 Garching, Germany, and Steward Observatory, University of Arizona, Tucson, AZ 85721}
\author{and Donald P. Schneider}
\affil{Department of Astronomy and Astrophysics, The Pennsylvania State University,
 University Park, PA 16802}

\begin{abstract}
We present UV ($\sim 2300$ \AA) images, obtained with the
{\it Hubble Space Telescope} ({\it HST}) Faint Object Camera,
 of the central $20''$ of
 five galaxies containing circumnuclear star-forming rings.
The five galaxies are from a well-defined sample of 103 normal, nearby galaxies
we have observed with {\it HST}.
At the {\it HST} resolution ($0.05''$), the rings  break up
into discrete star-forming clumps. Each clump is composed
of many luminous ($L_{\lambda} (2300{\rm  \AA})\approx 10^{35}-10^{37}$ erg s$^{-1}$  \AA$^{-1}$)
and compact ($R\ltorder5$ pc) star clusters. These objects are similar to those that
have been recently reported in colliding and
starburst galaxies, and in several other circumnuclear rings.
A large fraction, 15\%--50\%, of the UV emission originates in 
these compact clusters. Compact clusters therefore may be the preferred
mode of star formation in starburst environments.

 For one galaxy, NGC 2997,
we measure the UV-optical colors of the individual clusters using an archival
{\it HST} WFPC2 image at $\sim 6000$ \AA. Comparing the colors and luminosities to
starburst population synthesis models, we show that the clusters are less than 100 Myr
old and have masses of at least a few $10^3 M_{\odot}$, with some as high
as $10^5 M_{\odot}$. The UV extinction to those clusters that are detected in the UV
is at most a factor of 10. In NGC 2997,
the limits on the masses and the ages of the young clusters indicate that
 these objects will remain bound and evolve into globular clusters.
However, data in additional wavebands are needed to critically test this hypothesis.

The luminosity function of the clusters in the rings
is similar in shape to those measured for super star clusters
in other star-forming galaxies, and extends to luminosities lower by
several orders of magnitude.
All five of the UV-detected circumnuclear rings occur in barred or weakly barred
 spiral galaxies of type Sc or earlier. None of the five rings has an active
nucleus at its center, arguing against a direct correspondence between
circumnuclear star formation and nuclear activity.

\end{abstract}

\keywords{galaxies: star clusters  -- galaxies: starburst -- galaxies: nuclei}

\centerline{\it To Appear in the Astronomical Journal, June 1996}

\section{Introduction}

One of the most intriguing results of galaxy imaging with the
high angular resolution of the
{\it Hubble Space Telecope} {\it (HST)}
has been the discovery of extremely luminous, compact, young star
clusters in a variety of starburst environments. These include
the cooling-flow/merger-remnant galaxy NGC 1275
(Holtzman et al.\ 1992), merging
galaxies (Whitmore et al.\ 1993; Conti \& Vacca 1994; Whitmore \&
Schweizer 1995), ``amorphous'' peculiar galaxies (Hunter, O'Connell,
\& Gallagher 1994;
O'Connell, Gallagher, \& Hunter 1994; O'Connell et al. 1995), 
 two  barred galaxies with circumnuclear rings (Barth et al. 1995),
 and nine starburst galaxies with a range
of morphologies and luminosities (Meurer et al. 1995).
  Prior to {\it HST\/}, only a few objects of
this type were known to exist (e.g., Arp
\& Sandage 1985; Melnick, Moles, \& Terlevich 1985); the severe crowding in most
starbursts prevents their resolution into individual clusters in
ground-based images. Similar objects have been detected recently 
in high-resolution infrared images of the starburst galaxy
NGC 1808 (Tacconi-Garman, Sternberg, \& Eckart 1996).

  Such ``super star clusters,'' having diameters
not exceeding a few tens of parsecs and luminosities greater than $M_V =
-10$ mag, result from extreme episodes of rapid star formation which are
virtually absent in ordinary galactic disks.  The small radii, high
luminosities, and presumably high masses of these clusters have led
to suggestions that, as opposed to standard OB associations,
 they may remain bound systems.
  If so, after $10^{10}$ years they would have
magnitudes comparable to those of old Galactic globular clusters, and therefore
could be present-day versions of young globular clusters. Van den Bergh
(1995) has argued that the sizes and luminosity functions
 of the young clusters more
closely resemble those of Galactic open clusters than those of globular clusters.
Meurer (1995b) counters that the sizes of
the young clusters are often poorly constrained in the {\it HST} data,
and that the luminosity functions of the globular clusters and
the super star clusters should not be directly compared because the
super star clusters in a given galaxy are not all necessarily of the same age
and so 
are viewed at different epochs (and luminosities) in the clusters' evolution.
 Note that luminous, compact clusters
are also known in our Galaxy and the Large Magellanic
Cloud (LMC). For example, recent {\it HST} and near-infrared observations
of the R136 cluster in 30 Doradus in the LMC indicate a mass of $\sim 10^4 M_{\odot}$
within a radius of 2.5 pc (Brandl et al. 1996). The dynamical crossing time
for such a cluster is $3\times 10^5$ yr, while the measurements constrain
the stellar ages to 3--5 Myr. This system is therefore bound, and could also evolve
into a low-mass globular cluster. The limits on luminosity used to define
super star clusters and distinguish them from the more common
and slightly less-luminous compact
clusters are arbitrary; physically, these are all similar objects. 
Clearly, learning more about these young clusters can shed light on
rapid star formation, globular-cluster formation, and the
possible connections between them.

The bulges of some nearby spiral galaxies display
intense star formation in a ring of optical
``hot-spots'' (S\'ersic \& Pastoriza 1967). Such galaxies
 offer us an opportunity to
study starburst properties at a level of detail that would be
impossible for more distant or more heavily obscured systems.
Kiloparsec-sized nuclear rings can form as a result of bar-driven
inflow to an inner Lindblad resonance (e.g., Tubbs 1982; Athanassoula
1992; Piner et al.\ 1995), yielding a large concentration of dense gas
in a small region surrounding the nucleus (Kenney et al.\ 1992).
Several nuclear rings
have been studied extensively from the ground (e.g., Hummel, van der Hulst,
\& Keel
1987), but the small sizes of the rings (typically $\sim10$\arcsec)
have precluded detailed studies of their structure.  {\it HST\/}
imaging has opened a new vista on these hot-spot galaxies.  WFPC,
WFPC2, and FOC images have revealed large populations of super star
clusters in nuclear rings in four nearby galaxies: NGC 4314 (Benedict et al. 1993),
 NGC 1097 and NGC 6951 (Barth et al. 1995), and NGC 3310 (Meurer et al. 1995).

The process of cluster formation in these starbursts appears to be
physically distinct from normal star formation in spiral galaxies.
Even the
most luminous disk \ion{H}{2} regions in late-type galaxies seldom
contain compact clusters that might be potential young globulars
(Kennicutt \& Chu 1988). The occurrence of super star clusters may be
unique to starburst environments.
Hot-spot \ion{H}{2} regions are also spectroscopically distinct from
disk \ion{H}{2} regions; the hot-spot \ion{H}{2} regions have higher
heavy-element abundances and possess lower emission-line equivalent
widths than disk \ion{H}{2} regions of the same luminosity, suggesting
that continuous, intense star formation occurs within the hot-spots
(Kennicutt, Keel, \& Blaha 1989; Mayya 1995).  Circumnuclear hot-spots
contain over 50\% of the massive star formation in some galaxies
(Phillips 1993); thus, determining the physical parameters of hot-spot
\ion{H}{2} regions is important for understanding the global star
formation properties of spirals.

We have carried out a UV imaging survey with the {\it HST} Faint
Object Camera (FOC) of the central regions of 103 large, nearby 
galaxies, selected randomly from a complete sample of 240 galaxies.
An atlas of the survey is presented
by Maoz et al. (1996), where full details of the sample selection,
 observations, and data reduction are given.
 One of the main purposes of the survey
is to study star formation in the central regions of the galaxies. By
observing at $\sim2300$ \AA, the survey images detect predominantly 
the youngest existing stellar populations and thus provide a clean probe of the
most recent sites of active star formation, uncontaminated by light
from more evolved stars.

In this paper, we present results for five out of the 103 galaxies, whose
UV images display circumnuclear star-forming rings. In a forthcoming
paper (Ho et al. 1996c) we will analyze star formation based on the UV
images of the other sample galaxies.
In \S 2, below, we describe the observations and data analysis. Notes on the individual
galaxies
are given in \S 3. We discuss the physical implications of the data in \S 4
and summarize our results in \S 5.

\section{Observations and Analysis}

The five galaxies described in this paper were observed with {\it HST} in
1993, June and July. Here we briefly outline
the sample from which the five galaxies are taken, the
 observations, and the data reduction. More complete details are given
in Maoz et al. (1996).

 The sample consists of all
galaxies in the UGC and ESO catalogs (Lauberts \& Valentijn 1989)
 with heliocentric velocities
less than $2000$ km s$^{-1}$ and photographic diameters (as 
defined in the catalogs) greater than $6'$.
Exclusion of 22 galaxies for a number of reasons (see Maoz et al. 1996)
left 240 galaxies from which the Space Telescope Science Institute (STScI)
staff chose Snapshot targets based only on scheduling convenience.  
A total of 103 out of the 240 sample galaxies were successfully observed
while the program was active. 

Data were obtained with the {\it HST f/}96 Faint Object Camera (FOC; Paresce 1990)
in its ``zoomed'' $1024\times 512$-pixel mode
 with $0.022''\times 0.044''$ pixels,
 giving a field
of view of $22''\times 22''$. The F220W filter was used. This is
a broad-band filter with an effective wavelength of $\sim 2300$  \AA\ and
effective bandpass of $\sim 500$  \AA. 
 The exposure time was 10 minutes per galaxy.
The images were processed by STScI's ``pipeline'' reduction (Baxter
et al. 1994), after which the pixel scale is $0.0225''$ pixel$^{-1}$. All our data
were obtained before the {\it HST} repair mission at the end of 1993,
and therefore are affected by spherical aberration. 
As a result, the point-spread function (PSF) consists of a sharp core
of full width at half maximum (FWHM) $\sim 0.05''$ that contains about 15\% of the light,
with the rest of the light
 spread in a complex low-level ``halo'' with a radius of several arcseconds 
(Burrows et al. 1991).
In the observing mode we have used, the FOC is limited in its dynamic range
to 255 counts (8 bits) per zoomed pixel; additional signal causes the
counts to ``fold over'' and start again from 0. Another problem
is that the detected count rate becomes nonlinear, gradually saturating
 for bright sources (see Baxter et al. 1994). 
The central pixels of most of the compact bright sources detected in the 
images may be in the nonlinear regime, and the brightest of them
are clearly saturated. Our analysis will therefore rely mainly
on the wings of the PSF, which have low count rates
 ($\ltorder 0.05$ s$^{-1}$ pixel$^{-1}$). We 
model the PSF  using a well-exposed
F220W image of a star observed with the FOC $ f/96$ $256\times256$ pixel mode,
which has a large
dynamic and linear range (but small field of view). Such empirical PSFs
are required for work in the UV (Baxter et al. 1994).

   As in Maoz et al. (1995, 1996), we translate the FOC counts to a flux density
   $f_{\lambda}$ at 2270 \AA\ assuming
    1 count s$^{-1}=
   1.66\times 10^{-17}$ erg s$^{-1}$ cm$^{-2}$ \AA$^{-1}$, based on
   the on-line calibration data available from STScI for the FOC and F220W
   filter, with a 25\% increase
   in sensitivity of the $512 \times 1024$ zoomed-pixel 
   mode relative to the $ 512\times 512$ pixel
   mode (Baxter et al. 1994). This calibration assumes a spectrum that is
constant in $f_{\lambda}$. As detailed in Maoz et al. (1995), the F220W count-rate
vs. $f_{\lambda}$ at 2270 \AA\ is  weakly dependent on the spectral slope,
with a change of only a few percent for a large range in slopes.

 The uncertainty in the absolute flux is $\sim 20$\% when measuring
 individual compact sources (Baxter et al. 1994), but can be as small
 as $\sim 5$\% when measuring the UV flux integrated over large areas
 (e.g.\ $\sim 150$ arcsec$^2$; Meurer 1995a), if the background can be
 reliably determined.

In the analysis below, we also use a pair of optical images of NGC 2997 taken
on 1994 June 21 (after the repair mission) with the {\it HST} WFPC2, and obtained
from the {\it HST} archive. They are two
80 s exposures through the F606W filter (similar to $V$-band).
The star-forming ring of interest is on the PC detector,
 which has a pixel scale of $0.045''$ pixel$^{-1}$. The images were pipeline
processed, registered, and co-added to reject cosmic-ray hits.
Counts were converted to flux at 6060 \AA\ using the STScI pipeline
calibration and assuming a flat spectrum in $f_{\lambda}$.

The FOC images show a variety of morphologies and UV 
brightnesses in the centers of the galaxies (Maoz et al. 1996).
Maoz et al. (1995) analyzed the data for nine galaxies whose images
show conspicuous bright and compact sources at the galaxy nucleus.
In this paper, we present results for five out of the 103 galaxies, whose
UV images display circumnuclear star-forming rings. The five were found
by visually examining all the images for any UV-emission morphology 
roughly surrounding the nucleus (i.e., the center of the image) in a ring,
or in patches roughly forming a ring. Four of the five galaxies selected
in this way are already known  from previous ground-based images
to be galaxies with circumnuclear star-forming rings
 (NGC 1433, NGC 1512, NGC 2997, NGC 5248; see \S 3).
One of the five, NGC 1079,
was not previously known to be a circumnuclear-ring galaxy, probably
due to its small ($4''$) ring diameter. As we show below, the physical size
and properties of its ring are similar to those of the other four galaxies.
Figure 1 displays the FOC UV images of four of the rings. Figure 2 shows
the FOC (UV) and WFPC2 (optical) images of NGC 2997, shown with matched
orientation and scale.  Table 1 lists the
main characteristics of the ring galaxies.

As seen in Figures 1 and 2, the rings in the {\it HST} images break up into a number
of clumps, with each clump consisting of many compact sources on top of a 
diffuse background. The background is highly irregular and often cut
by dust lanes. (The detail afforded by the {\it HST} angular resolution
can be appreciated by comparing NGC 1512 in Fig. 1 to a ground-based
H$\alpha$ 
image in Fig. 2 of Kennicutt [1994].) Note that not only is the area exterior to the rings
mostly devoid of UV emission, but so is the region interior to the rings,
including the nuclei of the galaxies.
 We have matched up the gross features of the rings with
optical ground-based images (and in the case of NGC 2997, with an optical
{\it HST} image; see \S 3.4 and Fig. 2), and verified that the nucleus of the galaxy
is indeed within the ring but invisible in the UV (except for NGC 1512,
which has a faint nuclear UV source). This shows that, despite the high
nuclear surface brightnesses, little visual-band light leaks through the
F220W filter (see also Maoz et al. 1996),
 and mostly the youngest stellar populations are registered in the images.
(The ``UV-upturn'' observed in some elliptical galaxies and some spiral-galaxy bulges
[e.g., Burstein et al. 1988],
and thought to originate from evolved stars, sets in at $\lambda \ltorder 2000$ \AA,
and may contribute some of the diffuse UV emission we observe.)
We list in Table 1 the total 2270 \AA\ flux from each of the rings. We measured
it by summing the counts above the background, which was determined in the empty corners
of each image. The main background source at $\sim 2300$ \AA\ is particle induced,
and variable throughout the {\it HST} orbit (Nota, Jedrzejewski, \& Hack 1995).
 Since the galaxies observed are
much larger than the field of view, we cannot exclude some contribution to the background
from diffuse UV emission from the galaxies themselves.
The UV fluxes quoted in this paper are uncorrected for Galactic or intrinsic extinction.
Galactic extinction is negligible for the galaxies discussed here, except
for NGC 2997, which has $E(B-V)=0.1$ mag (Burstein \& Heiles 1982; see also \S 4.2),
corresponding to a UV extinction $A_{2300}\approx 0.8$ mag.

The compact sources distributed along the rings are reminiscent of the 
super star clusters detected in optical and UV {\it HST} images of other 
starburst galaxies (see \S 1). To study their properties, we have
measured the brightnesses and intrinsic sizes of the individual sources.
At least some of the diffuse emission surrounding the bright
compact sources is an artifact of the spherically-aberrated PSF. By summing the
contribution of the individual sources to the total UV light, we can
estimate what fraction of the active star formation in the rings is occuring
within these compact objects.

To identify all compact sources in each image in a fairly objective way,
the DAOFIND algorithm in the IRAF \footnote{IRAF (Image Reduction and Analysis
Facility) is distributed by the National Optical Astronomy Observatories,
which are operated by Aura, Inc., under cooperative agreement with the
National Science Foundation.} implementation of DAOPHOT (Stetson 1987)
was applied to the data. The threshold for detection of a source above
 the background was set
low enough so that most of the objects clearly discerned by visual inspection would
be detected, but high enough to avoid an excessively large number of false identifications
of artifacts and noise peaks. A fair number of sources was always missed by
DAOFIND, due to the complicated background, and a number of false identifications
made, e.g., at the intersection of PSF rings. The detected objects were therefore
 marked on a screen,
obvious misidentifications were deleted, and obvious sources that were missed
by the algorithm were added to the list of detected sources. Although this process
is not completely objective, it is the best compromise we have found for
these data.

As already noted, the central pixels of every compact source are suspect of
nonlinearity and/or saturation, and so cannot be used in a conventional
PSF-fitting-based crowded-field photometry program.
Maoz et al.
(1995) fitted empirical PSFs to the azimuthally-averaged radial profiles of bright nuclear
sources from this survey, between radii of $0.4''$ and $2.2''$ (i.e. the ``rings'' and ``halo'' of the
PSF), excluding the central, saturated pixels.
 The sources detected in the ring galaxies discussed here
 are significantly fainter
than those nuclear sources and are also crowded, so this technique is not applicable.
 Instead, we follow Meurer et al. (1995) and fit the sources with the
radial profile calculated in the wings of an empirical PSF, at radii of
$0.044''$ to $0.2''$ (i.e., 2 to 10 pixels, again excluding the central, high count-rate pixels).
For the brightest compact sources in the rings, where the PSF wings are also 
saturated, we use the Maoz et al. (1995) algorithm. In either case, the PSF is
convolved with Gaussians of varying widths, and its radial profile is scaled in
brightness and is added to a constant background so as to minimize the $\chi^2$ fit
to the radial profile of each source. The convolved Gaussian width of the best-fit
PSF provides an estimate of the intrinsic size of the sources. 

To gauge the realiability of the size and brightness measurements in the
presence of noise, crowding, and variable background, we
have carried out the following experiments. First, we added to the images simulated
sources of known brightness and size on regions with different
backgrounds, added the appropriate Poisson noise, and then measured the sources.
Second (as in Meurer et al. 1995),
 we tested the algorithm by applying it to the stars in F220W FOC images of the Galactic
cluster NGC 104. From these two tests we conclude that compact-source brightnesses can
be measured to $\sim 20\%$ accuracy for sources with $> 1$ count pixel$^{-1}$ at the
2-pixel radius in
their radial profiles. Such a brightness corresponds to 250 total reconstructed counts
in the source, or $7\times 10^{-18}$
 erg s$^{-1}$ cm$^{-2}$ \AA$^{-1}$ in these 600 s exposures. We have tabulated in this paper
only sources whose total reconstructed counts are greater than 100, i.e.
 $f_{\lambda}(2270{\rm \AA}) > 3\times 10^{-18}$
 erg s$^{-1}$ cm$^{-2}$ \AA$^{-1}$. This can be considered the detection limit of these images
for relatively isolated sources on a low background. The detection limit is higher
in crowded, high surface-brightness regions.
 Recovering the width of the Gaussian that was 
convolved with the PSF is more difficult, and is reliable to any degree only for
sources with brightness $> 4$ counts pixel$^{-1}$ at the
2-pixel radius (i.e. $\gtorder 3\times 10^{-17}$
 erg s$^{-1}$ cm$^{-2}$ \AA$^{-1}$). The resolution limit of the technique
 is reached for a Gaussian
$\sigma$ of 1.5 pixels (or $0.033''$, i.e. objects with an intrinsic
 Gaussian FWHM of $< 0.08''$ appear unresolved).
 Recovered widths are accurate to about 0.5 pixels.
 Sources wider than 4 pixels are often elongated and faint, and are probably 
composed of several poorly resolved clusters.
Their width is also poorly reconstructed by the technique, which assumes radial
symmetry.
The width estimate can be
led completely astray by crowding or a high and variable background (due to, e.g., the
PSF features of another bright source). All width estimates given below should therefore
be treated with caution until confirmed by higher $S/N$, unaberrated images.
The true width is, nevertheless, unlikely to be much larger than in our estimates,
since it is difficult to make a source appear more compact than it really is.

The brightnesses of the compact sources in the optical WFPC2 image of NGC 2997
were measured using DAOPHOT aperture photometry, as described in
Barth et al. (1995) for the measurement of compact sources in WFPC and WFPC2
optical images of two other circumnuclear ring galaxies, NGC 1097 and NGC 6951.
The photometry used a 2-pixel radius aperture, the background was determined
in an annulus of radius 4-6 pixels, and an aperture correction was determined using
an artificial PSF (created with the Tiny Tim package; Krist 1994),
broadened by convolution with a Gaussian to match the mean width of
several of the brightest clusters in the image. 
 Due to the inconstant background and the undersampling
of the PSF, the typical errors are of order 10\%.
We tabulate from the optical image objects with 
 $f_{\lambda}(6060{\rm \AA}) \ge 1\times 10^{-18}$
 erg s$^{-1}$ cm$^{-2}$ \AA$^{-1}$, which at the distance of
NGC 2997 corresponds to $M_V \le -7$, i.e. brighter than
the most luminous individual stars.

Tables 2-6 list the results of the measurements of the individual compact sources
for each of the galaxies. The geometrical center was estimated for each ring
and is listed in the first line of each table (source No. 0). The offsets of
all the sources are listed relative to this coordinate origin.
 The best-fitting widths are listed only for those sources
that are bright and uncrowded enough to make this measurement fairly reliable.
 For NGC 2997, we have matched those sources appearing in both the UV and optical
images, and list them as such in Table 5.
		
\section{Notes on Individual Objects}
\subsection{NGC 1079}
As seen in Figure 1, the ring in this SAB0 galaxy is the smallest in angular extent
($4''$) which we have detected. This probably explains why it has not been
noted before from ground-based observations. At a distance of $16.9 h_{75}^{-1}$ Mpc
(Tully 1988), the major-axis
 diameter of the ring is $330 h_{75}^{-1}$ pc. To verify that it is, in fact,
a star-forming ring (and not, e.g., an edge-brightened ``bubble''), we obtained a long-slit
spectrum along the ring's major axis. The observations were done 
on 1993 November 8 UT at the Lick 3-m 
Shane reflector using the Kast Cassegrain spectrograph (Miller \& Stone 
1993). Two exposures, each of half hour duration, were obtained
with a slit width of $1^{\prime\prime}$, giving a resolution of 
$\sim 2.3$~\AA\ over the wavelength range 4550 \AA -- 5260 \AA.

We used stellar and galaxy templates and a least-squares
fitting process (Rix \& White 1992; Rix et al. 1995) to derive rotation curves along the
major axis both for the stars (from stellar absorption features) and for the
gas (from emission features).
The results (Figure 3) show within $\pm 2''$ (the radius
of the ring) the linear rise
expected for a ring in projection. Assuming the ring is circular, the axis ratio ($\sim 2$)
implies an inclination of $60^{\circ}$, and a deprojected circular velocity
at $2''$-radius of $\sim 60$ km s$^{-1}$. The velocity implies a mass interior to the
ring of $10^8 M_{\odot}$, and a mean mass density of $15 M_{\odot}$ pc$^{-3}$ within
the central 160 pc, normal (though somewhat on the low side)
for the centers of galaxies.
 The emission-line spectrum is typical of a starburst.

Similar to the case in some other galaxy bulges (e.g.,
Fillmore, Boroson, \& Dressler 1986;
Rix et al. 1995),
the measured rotation velocity of the stars is lower than that
of the gas. This may have two reasons: First, the stars have a
larger degree of support from random motions and hence their
streaming motion is smaller. Second, the observed rotation
includes an integration along the line-of-sight. If the
ionized gas is more centrally concentrated and more flattened
than the stars, its rotation will be less affected by this 
integration.

We identify 19 compact sources in the {\it HST} UV image,
 whose brightnesses are listed in Table 2.
Comparing their summed contribution (including the PSF halos)
 to the total UV light measured in the image,
we find that 23\% of the UV light is contributed by discrete compact sources.
In its western half, the ring is either broken or transversed by a dust lane.

\subsection{NGC 1433}
This SBab galaxy has been noted, apart from its nuclear ring, for several outer rings
(Kinney et al. 1993, and references therein). At the angular resolution of the {\it HST}
image, the ring appears irregular and patchy. Interestingly, the nuclear ring in
NGC 1512 is much more ordered, even though both galaxies have very similar sizes,
luminosities, and morphologies (see Sandage \& Bedke 1988, for beautiful pictures of these
two galaxies side by side). From comparison of the UV image of NGC 1433 to a ground-based
CCD image through a narrow-band H$\alpha$ filter, we verify that the nucleus of the
galaxy is within the UV ring structure, but is dark in the UV. 

The major-axis diameter of the ring is $\sim 9.5''$, or $540h_{75}^{-1}$ pc at a distance
of $11.6 h_{75}^{-1}$ Mpc (Tully 1988). We identify 31 compact sources in the image, which
contribute 12\% of the total UV light.

\subsection{NGC 1512}
The spiral ring in this SBab galaxy is highly ordered and reminiscent of that seen
in NGC 6951 (Barth et al. 1995). The ring major axis is $\sim 15''$,
 $690h_{75}^{-1}$ pc at a distance of $9.5 h_{75}^{-1}$ Mpc (Tully 1988).

We detect 43 compact sources in the image, including one (No. 31) on the southeast side of the 
ring whose brightness is 1.5 times greater than all the others combined.
There is
also a faint compact source at or near the nucleus (No. 12). 39\% of the UV light is in
the detected compact sources, or 16\% excluding the brightest source.

\subsection{NGC 2997}

The nuclear ring of this SABc galaxy has been previously observed from the ground
(Meaburn \& Terret 1982; Walsh et al. 1986). As in NGC 1433, the ring in the HST
UV image is patchy and irregular, and very faint along its southern half.
The major-axis diameter is $\sim 8''$, or $540 h_{75}^{-1}$ pc at $13.8 h_{75}^{-1}$ Mpc.
The 39 compact sources we detect contribute 14\% of the total UV light.

Fortunately, we have found in the {\it HST} archive a WFPC2 F606W (i.e. $\sim 6060$ \AA)
exposure of this galaxy (see \S 2). The 155 compact sources we measure in the optical
image are listed in Table 5. By computing the coordinate transformation between the
UV and optical images, we have been able to identify 24 of the UV-detected sources
in the optical image.
The UV-optical colors we can extract for individual compact sources in this galaxy provide 
important information on the physical nature of the sources. We exploit these
data further in \S 4.2.
There are 133 optical sources that are undetected
in the UV image.  The brighter among the optical-only sources may be either
aged or highly-reddened star clusters.
 The fainter of these may also be smeared by the spherical aberration
in the UV image into some of the diffuse emission observed there. We examine the
potential importance of this effect in \S 4.1. The faintest among the optical-only sources
 could also be individual
red luminous stars, such as supergiants. The 15 sources (mostly faint) that are detected
only in the UV must be either faint young clusters or OB associations,
or individual young massive stars. There is no obvious difference in the spatial
distributions of the clusters that are detected in either one or both bands.

\subsection{NGC 5248}
This SABbc galaxy has a well-ordered
nuclear ring. The major-axis diameter is $\sim 11''$, or $1.2 h_{75}^{-1}$ kpc
at $22.7 h_{75}^{-1}$ Mpc. The 46 compact sources we detect contribute 16\%
of the UV light. The nucleus, which is invisible in the UV, has the optical
spectrum of an H II region through a $2''\times 4''$ aperture at position angle
$45^{\circ}$ (Ho, Filippenko, \& Sargent 1995, 1996a). However, 
this spectrum might have a significant
contribution from the H II regions in the ring.

\section{Discussion}

\subsection{The Fraction of Light in Compact Sources}

The UV luminosities (uncorrected for reddening)  of the
compact sources distributed in the rings, listed in Tables 2-6,
 are in the range $L_{\lambda}(2270{\rm  \AA})\approx 10^{34.5}-10^{37.5}$
erg s$^{-1}$  \AA$^{-1}$, where $L_{\lambda}(2270{\rm  \AA})
\equiv f_{\lambda}(2270{\rm  \AA}) 4\pi D^2$ and $D$ is the distance
to a galaxy. For reference, $\log L_{\lambda}(2270{\rm  \AA})$
 is 34.4 and 31.2 for a single main-sequence
B0 or A0 star, respectively (Allen 1973).
The optical luminosities of the objects in NGC 2997,
listed in Table 5, are in the range
 $L_{\lambda}(6060{\rm  \AA})\approx 10^{34.3}-10^{36.2}$
erg s$^{-1}$  \AA$^{-1}$. Also for reference, $M_V=-10$ mag,
used by O'Connell et al. (1994) as a defining limit for super star clusters,
corresponds to  $\log L_{\lambda}(6060{\rm  \AA})\approx 35.6$.
Those sources bright
enough for a size measurement typically have a Gaussian radius
of only several pc, and are often unresolved, with upper limits
of $\ltorder 2$ pc on their radius.
The brighter among these objects 
are in all respects similar to the
``super star clusters'' or ``young globular clusters''
identified in {\it HST} images of other starburst environments
(see \S 1).\footnote{ Note that the high luminosities of the objects
in the galaxy He 2-10, observed with the same observational
setup as ours, and reported by Conti \& Vacca (1994), result
from applying an uncertain extinction correction of a 
factor of $\sim100 $
to the UV fluxes. The uncorrected UV luminosities are 
in the range $\log L_{\lambda}(2270{\rm  \AA}) \sim $
 35.5 to 36.5, similar to that of the compact sources reported here.}
It is becoming clear that these compact sources, whatever
their nature, are quite common in starburst environments.
In the absence of color information, the fainter objects
in the UV images could be individual O and B stars.
In the optical image of NGC 2297, the faintest sources have
luminosities comparable to those of the most luminous red supergiants
 ($M_V\approx-7$; Allen 1973) and so could also be individual stars.

As shown in \S 3, a substantial fraction of the UV light, ranging from 
10\% to 40\% among the
five rings, is coming from clearly detected sources. Similar
fractions were reported in the starburst galaxies analyzed
 by Meurer et al. (1995). Since  the UV images are sensitive mostly to
the youngest stellar populations, it appears that
these compact clusters are one of the preferred modes of 
star formation. Furthermore, the fractions given above are lower
limits to the true fraction of UV light coming from individual sources,
because the overlapping PSF halos of faint objects that are below our detection limits
for individual sources will contribute
to the diffuse component observed in an image. To examine the potential
importance of such an effect, we have taken advantage of the optical
WFPC2 image of NGC 2997. The optical image reveals many more sources,
especially faint ones, than the FOC UV image. This could be 
because of the better resolution (and hence sensitivity to faint objects)
of the un-aberrated optical image, or because the fainter sources
are intrinsically redder (due, e.g., to age or reddening of a cluster)
 and hence invisible in the UV, or a combination of both. 

To study this question we
 performed the following simulation. We convolved the WFPC2 optical image
with the spherically-aberrated PSF, and scaled the resulting image
so that 10 of the medium-brightness clusters would have similar counts to those
measured in the FOC UV image. After subtracting a constant to make the
background similar to that measured in the FOC image, we added Poisson
noise, and so obtained a simulated FOC image of how the galaxy would
appear if all the compact sources had a UV-optical color similar to that of the
medium-bright clusters. We find that, in fact, the simulated FOC image
is similar to the actual observed one, with most of the faint optical
sources undetectable.
 This indicates that, at least in this galaxy, the
actual fraction of the UV light coming from clusters could be
larger than measured (14\%), simply due to the lower sensitivity of
the UV setup. Since 60\% of the light from compact sources in the optical
image is from sources that are undetected in the UV image, the true
fraction of the UV light contributed by compact sources could be as high 
as $2.5\times 14\%=35\%$. If such a factor applies to the other galaxies
as well, the corrected fraction there will be as high as $\sim 50\%$.
Alternatively, the assumption of similar color for all clusters may be 
wrong, and the faint objects may be undetected in the UV because they
are too red. If so, the measured UV fraction in clusters is representative
of the true fraction. This question can be resolved by deeper UV observations
of these galaxies with the now-repaired {\it HST} optics. In any case,
the contribution of the clusters to the UV light is substantial, reaching
nearly one-half in NGC 1512.

\subsection{The Nature of the Compact Sources}
In the previous studies that have reported UV observations of super star clusters,
or observations of clusters in nuclear rings, data at the
{\it HST} angular resolution were obtained in only one broad band per galaxy.
Calculations of the extinction-corrected luminosity and color
of the clusters depended on extinction factors derived 
from ground-based imaging and spectroscopy having much lower angular resolution
than the {\it HST} data. These corrections are uncertain in
view of the heterogeneous mixture of stars, clusters, and dust seen
in the {\it HST} images. It is also conceivable that the detected clusters
are selectively those that are least obscured. In the absence of reliably de-reddened
luminosity and color information, a discussion of the masses and ages,
and hence nature of the clusters, is speculative.
While the data for four of the five galaxies presented here suffer from
the same single-band problem, the UV and optical images of NGC 2997 offer 
us the opportunity for the most detailed examination to date of the nature
of the clusters.

Using the fluxes in Table 5, we list in Table 7
 for every compact source in NGC 2997 that is detected in both the
optical and the UV images a 6000 \AA\ -- 2300 \AA\ 
color and an observed luminosity.
We relate these observables to the reddening, mass, and age of each cluster
by calculating stellar cluster synthesis and evolution models, as described
by Krabbe et al. (1994) and by Kovo \& Sternberg (1996).
 Briefly, we assume an initial mass function (IMF),
a star formation rate, and its functional dependence on time (e.g.,  a constant, continuous
rate of star formation, or an exponentially decreasing burst with some characteristic
decay time). The evolution of the
 luminosity and broad-band spectrum of the different stellar types are traced using
 solar-metallicity Geneva stellar tracks (Schaller et al. 1992). The broad-band properties
are weakly dependent on metallicity (e.g., Leitherer \& Heckman 1995).
 For any input IMF (including slope, and
upper and lower mass cutoffs), star formation rate, and cluster age, the integrated
 luminosity at any wavelength and the total stellar mass can be computed.
 ``Color-color'' and ``color-magnitude''
diagrams can thus be produced, the effects of reddening by a chosen
extinction law can be applied to the calculation, and the results 
compared to the observations. 

Our models are similar to those presented in the literature by
several investigators
(e.g., Larson \& Tinsley 1978; Mas Hesse \& Kunth 1991; Bruzual \& Charlot 1993;
Vacca 1994; Leitherer \& Heckman 1995).  
We have compared our model results with those of Leitherer \& Heckman (1995)
and find good agreement.  For example, our computed V magnitudes agree
well (generally to within 0.3 magnitudes)
with those of Leitherer \& Heckman for both continuous and decaying
bursts and a range of IMF slopes and upper mass cut-offs.

Figure 4 shows a ``color-magnitude'' diagram resulting from such a calculation,
relating the 6000 \AA\ luminosity to the 6000 \AA\ -- 2300 \AA\ color. The two trajectories
show how two clusters of two different total masses move on the diagram with
time (the total mass simply scales with the luminosity). We have assumed an
IMF with a power-law slope of $-2.5$,
 upper and lower stellar mass cutoffs of $30 M_{\odot}$
and $1M_{\odot}$, and an exponentially decreasing star formation rate with characteristic
time 5 Myr. The total cluster mass is that which is reached asymptotically.
The points plotted in Figure 4 give the color and luminosity of all the NGC 2997 clusters
detected in both bands, as listed in Table 7.
 The arrow in Fig. 4 shows the de-reddening vector for
an extinction law
\footnote{Meurer et al. (1995) have studied in detail the effects of four different
Galactic and extragalactic
reddening laws on measurements through the FOC F220W filter. They show that 
three of the four give $A_{2300}\approx (8.25\pm0.25) E(B-V)$  for $0<E(B-V)<0.5$ mag,
as one would expect from a simplistic $A_{\lambda}\propto \lambda^{-1}$ model.}
 of the form $A_{\lambda}\propto \lambda^{-1}$ mag. A visual
 extinction $A_V=0.3$ is expected for 
NGC 2997 from foreground Galactic absorption alone (Burstein \& Heiles 1982).

Several inferences can be made from Figure 4, given the assumptions regarding the
IMF and the burst-like star formation rate.
 If the clusters are unreddened, their total
masses are between a few thousand and $10^5 M_{\odot}$, and their ages are 10 to 100
Myr. A de-reddening correction will move the points leftward (bluer) and upward (higher
luminosity) on the plot. Therefore, the ages of the clusters, as read off the plot,
are upper limits to the true ages. {\it All 24 clusters detected in both bands are younger
than $\sim 100$ Myr.} Similarly, the masses obtained from the plot are lower limits to
the true masses. {\it Most of the clusters have masses $\gtorder 10^4 M_{\odot}$}.
A de-reddening correction will move
the points by about half a decade in luminosity for every decade in color.
The reddest clusters can be moved leftward by about one decade, and most clusters
by only about half a decade, before they become bluer than a newly-formed cluster
of stars. The corollary is that {\it the clusters seen in both bands are extinguished by,
at most, about 1 mag in the visual}, and the mass estimates given above for
the unreddened case can be off by a factor of a few, at most.

The above conclusions suggest that the compact sources seen in NGC 2997
are indeed young clusters of stars. The typical radius of these objects
is $\ltorder 3$ pc. If there is little reddening, and the true ages
of the clusters are close to the upper limits (i.e. tens of Myrs), then 
the clusters appear to be bound -- otherwise they would not have remained
compact
for so long. For example, for an unbound cluster to retain a 3 pc radius
for 25 Myr, the velocity dispersion would have to be less than
 0.12 km s$^{-1}$. The density of the clusters is two orders of magnitude
greater than the galaxy density within the ring, so the clusters will
not be tidally disrupted. Our results are therefore consistent with
the idea that the clusters will evolve into objects similar to present-day 
globular clusters.

Do these conclusions hold if the clusters are reddened, and
hence very young? If a cluster of mass $M$ and radius $R$
 is unbound, its velocity dispersion $v$
is
$$
v > (2 G M / R)^{1/2},
$$
and so its maximal age is
$$
t_{max} < R/v < 3\times 10^4 {\rm yr} \left({R}\over{\rm pc}\right)^{3/2}
 \left({M}\over{10^5 M_{\odot}}\right)^{-1/2}.
$$
It is unlikely that all the clusters around the ring
are only a few tens of thousands of years old. Such instantaneous triggering
is difficult to envisage in a ring whose dynamical time scale is $\sim 10^7$ yr
(based, e.g., on the kinematics we have measured for the NGC 1079 ring, \S 3.1).
The clusters are therefore bound, even if they are reddened.

How sensitive are these conclusions to the initial conditions of the
population synthesis calculation? We find that the curves in the
color-magnitude diagram have the same
general shape for a star-forming-rate decay time of 1 Myr, rather
than 5 Myr, although the trajectories are slightly more jagged at early times 
due to variations in the post-main-sequence evolution of stars of different
masses. For completeness, Figure 5 shows the same kind of
calculation, but with a continuous, constant star-formation rate,
rather than an exponentially-decaying burst. The two curves show the
evolution of a cluster
 for two star formation rates, $5\times 10^{-4} M_{\odot}\ {\rm yr}^{-1}$
and  $5\times 10^{-5} M_{\odot}\ {\rm yr}^{-1}$.
 In this case, the total mass in
a cluster is just the star-formation rate times the age of the cluster,
and so is not constant. As can be seen, under the continuous
star-formation assumption the ages
of the clusters in NGC 2997 could be up to $\sim 500$ Myr, since the 
observed blue color is provided by young stars, which are constantly
produced in a cluster. However,
such a star-formation scenario is probably unrealistic, as it
requires stars to be formed only in select regions of radius 3 pc for 500 Myr
($\sim50$ orbital periods). The first supernovae would probably blow away much 
of the gas in such a small region, so continuous accretion from the surrounding
ISM is required. The continuous star-formation scenario is
less problematic if the clusters are reddened, and so can be moved in Figure 5
to lower ages.

We find the model calculations are only weakly affected if, instead of an IMF slope
of $-2.5$, we use a 
Salpeter (1955) slope of $-2.35$. This approximate slope has been observed
in a variety of star-forming environments
 (e.g., Lada et al. 1991; Lada \& Lada 1995).
The results are also weakly sensitive to the upper mass cutoff of the IMF,
since the most massive stars live for a very short time before exploding
and disappearing. Raising the lower mass cutoff would not change
the observables plotted in Figures 4-5, since the UV light is dominated
by massive main-sequence stars, while about half the optical light is contributed
by red giants that have evolved from intermediate-mass stars and the rest 
by massive main-sequence stars. The total
cluster mass,  most of which is in low-mass stars on the main sequence,
would increase by a factor of $\sim 2$ or decrease by a factor of $\sim 3$
 if the low-mass cutoff were at $0.1 M_{\odot}$ or $5 M_{\odot}$, respectively, instead
of at $1 M_{\odot}$. 

In fact, it has been claimed (Rieke et al. 1980, 1993) that the IMF in the
starburst galaxy M82 is deficient in low-mass stars. (This conclusion is,
however, controversial; see Scalo 1986.) Since much of the unobscured
star formation in M82 takes place in super star clusters (O'Connell et al. 1995),
they could be the site of this deficiency.
If the same mechanism operates in NGC 2997, 
its clusters would be different from present-day globulars, in
terms of mass function. The masses
and stellar content  of the ring clusters are still uncertain
and can be constrained
by observations at additional bands, which can 
disentangle the effects of reddening and age.

The range of UV luminosities of the clusters in the four other galaxies
presented here is similar
to that of NGC 2997. In the absence of information at other wavebands,
the masses and ages of the clusters are degenerate, even without the 
assumptions about the IMF, the star-formation history, and the extinction.
 From our models, we find that the  cluster mass dependence on the UV luminosity
and the age, after $t\gtorder 10^7$yr, can be parametrized with the following power laws.
For the model described above, with an exponentially-decaying star formation rate,
$$
M=3\left(\frac{L_{\lambda}(2270{\rm  \AA})}{10^{33}\ {\rm erg}\ {\rm s}^{-1}{\rm  \AA}^{-1}}\right)
\left(\frac{t}{10^7 {\rm yr}}\right)^{1.43}M_{\odot}.
$$
For the continuous star-forming model,
$$
M=2\left(\frac{L_{\lambda}(2270{\rm  \AA})}{10^{33}\ {\rm erg}\ {\rm s}^{-1}{\rm  \AA}^{-1}}\right)
\left(\frac{t}{10^7 {\rm yr}}\right)^{0.83}M_{\odot}.
$$
The UV properties of the other four galaxies are similar in all
respects to those for NGC 2997. It is therefore likely that the
conclusions regarding the age, mass, and reddening of the clusters in NGC 2997 will 
apply to those galaxies as well, once data in additional bands are 
obtained for them.

\subsection{The Cluster Luminosity Function}
Several of the previous {\it HST} studies that have detected super star clusters
have also measured their luminosity functions. Meurer et al. (1995) find that
the combined luminosity function $N(L)$ of the clusters they measure in most of their
starburst galaxies has a power-law shape, $N(L) dL \propto L^{\alpha} dL$,
 with slope $\alpha \approx -2$. Whitmore
\& Schweizer (1995) fit 
a power law of similar slope to the cluster luminosity function of the merging
system NGC 4038/4039. The luminosity function shown by Barth et al. (1995)
for the clusters in the circumnuclear ring in NGC 1097 is also consistent
with a power-law distribution of slope approximately $-2$.

It is interesting to see if the same luminosity function persists in
additional examples of clusters in the 
circumnuclear ring environment. Indeed, the luminosity function 
of the clusters in each of the five galaxies peaks near the detection limit,
and falls off at higher luminosities. However, the numbers are small.
Figure 6 shows the combined luminosity function for NGC 1433, NGC 1512, and
NGC 2997. These three galaxies are all at a similar distance ($\sim 10$ Mpc),
so the cluster detection limit is at a similar luminosity ($\log L_{\lambda}(2270{\rm  \AA})
\approx 34.5$). The detection limit should then
 not affect the shape of the combined luminosity function at luminosities above it.
Figure 6 compares the luminosity function to a power law of slope $-2$ (the dashed line,
which has slope $-1$ in logarithmic intervals of $L$). Within the limitations of the
small numbers,
 for $\log L_{\lambda}(2270{\rm  \AA})\ge 35.5$ the luminosity function
of the ring clusters appears to follow the same $\alpha=-2$ power law observed
for clusters in other environments and at other wavebands. Note that Meurer et al. (1995)
plot the cluster luminosity function only for clusters brighter than absolute
magnitude (as defined by them) $M_{220}=-14$. This magnitude, which is their defining
limit for super star clusters, corresponds to
 $\log L_{\lambda}(2270{\rm  \AA})>37.25$. Our result shows that this power law
shape  also describes the luminosity function at luminosities lower by
1-2 orders of magnitude. (However, the 
comparison is not straightforward, because Meurer et al. (1995)
have attempted to correct their cluster luminosities for extinction, whereas we have not.)

The $\alpha=-2$ power law overpredicts by a factor of 2 the number of
clusters in the $34.9\le \log L_{\lambda}(2270{\rm  \AA})\le 35.3$ bin.
A source at $D=10$ Mpc with $\log L_{\lambda}(2270{\rm  \AA})$ in this range produces
250-600 total counts in our FOC exposures. Such sources, although conspicuous
if isolated, can be difficult to
detect on a high background or in the neighborhood of brighter sources.
We suspect that the low number of clusters in this luminosity bin is the
result of this detection problem, rather than a real turnover in the 
luminosity function.

\subsection{The Occurrence of Circumnuclear Rings}

While there are some theoretical explanations for the occurrence of 
circumnuclear rings (see \S 1), and detailed modeling of individual
cases, there have been few statistics on the circumstances
under which these rings appear. Our sample provides
a good opportunity to study the statistics of nuclear ring
formation because\\
(1) the ringed galaxies were identified from an unbiased sample
drawn from a 
 complete,
diameter-selected galaxy sample;\\
(2) the {\it HST} resolution allows 
detection of small rings that would have been missed from the ground (e.g.
NGC 1079); and\\
(3) the UV bandpass emphasizes these star-forming structures
over the optically-bright but UV-dark central regions of a galaxy,
again aiding in detection.\\
On the other hand, the small field of view of the FOC will exclude
large, face-on nuclear rings in the more nearby galaxies, and dusty
nuclear rings will pass undetected in the UV.

To check the effectiveness of the UV images in detecting nuclear rings, we
have examined the FOC images of all the galaxies in our sample appearing
in the compilation of known nuclear rings by Buta \& Crocker (1993).
There are seven such galaxies. Four of the seven are the previously-known
ring galaxies discussed in this paper. Two of the seven (NGC 1543 and NGC 3486)
have nuclear rings larger than the FOC field of view, explaining why
we did not detect them as such. One of the seven (NGC 2903) is described
by Pogge (1989) as an ``incomplete'' nuclear ring. He shows an H$\alpha$
image of the galaxy, consisting mostly of an extended knot to one side of the nucleus.
The {\it HST} F220W image of the galaxy shows a large number
 of super star clusters whose geometry matches up with the gross features 
in Pogge's ground-based image. The geometry of the clusters is, however, too
random to be classified as a ring. This galaxy will be analyzed further
in Ho et al. (1996c). We conclude that our UV images are effective at detecting
nuclear rings when their angular scale fits on the FOC field of view.

Models have generally involved
the influence of bars in the formation of circumnuclear rings.
In the RC3 (de Vaucouleurs et al. 1991)
catalog, all five of the galaxies discussed here are classified
as barred (SB or SAB).
For comparison, 52 of the 103 
galaxies observed in the {\it HST} sample are classified as barred
by RC3. Our data therefore suggest that barred galaxies
are the preferred sites of nuclear ring formation. There are,
however, known examples of nuclear rings in unbarred galaxies,
such as NGC 7742 (Pogge \& Eskridge 1993; Wakamatsu et al. 1996).
Our statistics also suggest that at least 5/52, or about 10\%, of galaxies with
large-scale bars have a circumnuclear ring.

The five galaxies in the sample that have nuclear rings range in 
morphological type from S0 to Sc. Figure 7 shows the distribution of 
morphological types among the entire galaxy sample and the five 
ring galaxies, using the RC3 classification. There
is no clear preference for a particular morphological type, but there
is a trend toward earlier types.

\subsection{The Active Nucleus Connection}

The occurrence of circumnuclear rings may be connected to bar-driven
inflow. One of the outstanding problems in the theory
of active galactic nuclei (AGNs) is the mechanism that drives 
galactic material, with its large angular momentum, to the nucleus
and into the postulated massive black hole (see, e.g., Phinney 1994).
AGNs, bars, and circumnuclear
rings co-exist in some galaxies. It has been speculated that there
is perhaps an AGN in the center of every ring, possibly as the result
of the same infall-driving mechanism (e.g., Dultzin-Hacyan 1995).
On the other hand, a correlation could simply be the result of the
fact that rings tend to occur in early-type galaxies, which are the preferred
sites of LINER and Seyfert-type AGNs (e.g., Ho et al. 1996a). 
The absence of a bar-AGN connection 
is  demonstrated by Ho, Filippenko, \& Sargent (1996b),
who show
that AGNs are equally likely to occur in barred and unbarred galaxies.

Our sample allows a statistical  examination of the nuclear-ring/AGN connection. 
Among the low ionization
nuclear emission-line regions (LINERs; Heckman 1980)
 and Seyfert galaxies in the northern galaxies of the
 sample (Maoz et al. 1995; Maoz 1996), none has a circumnuclear ring
detected in the UV. Based on spectral classifications from the
literature, none of the five detected ring galaxies has a LINER
or Seyfert spectrum. The latter fact needs confirmation using
high resolution and high-S/N spectroscopy that may reveal faint AGN emission
lines on the stellar background, through apertures that will 
effectively exclude light from the rings. Since four of the five ring galaxies
are in the South, spectral classification of the nuclei of all the
southern galaxies in the {\it HST} sample would enable 
further testing for connections between rings and AGNs.
Tentatively, we conclude that, although examples of co-existing rings
and AGNs are known (e.g., NGC 1097 and NGC 6951; Barth et al. 1995), there is no
evidence for a clear correspondence between the two.

\section{Conclusions}

We have presented {\it HST} data for five nearby galaxies displaying
circumnuclear star-forming rings in the UV. The rings are composed of numerous
compact ($<$ few pc) and luminous objects, similar in all respects
to those that have been recently seen in other starburst environments
and presumed to be young star clusters.
One of the galaxies, NGC 2997, has an {\it HST} image in visual light, providing
color information for these objects at the {\it HST}
angular resolution.

We summarize our main conclusions as follows.\\
1. A substantial fraction (10\%--40\%) of the UV light in these galactic
circumnuclear regions
 is emitted by the compact sources. Considering our detection limits, the true
fraction may be larger, of order 30\%--50\%. Compact clusters
are therefore a common, and possibly dominant mode of star formation in circumnuclear
starbursts.\\
2. In NGC 2997, the typical compact source has the color of a late B to early A star
and the luminosity of $\sim 10^4-10^5$ such stars, or the luminosity of tens of OB stars.
The object must therefore be either a $\sim 30$ Myr-old unreddened cluster, or a young reddened
cluster. Either way, the mass of the typical object is $\gtorder 10^4 M_{\odot}$.\\
3. The blue color of the NGC 2997 clusters which are visible in both optical and UV bands
means that these clusters are not highly reddened, and are absorbed in the UV by less than 
a factor of 10.
 From the present data we cannot say whether the clusters that are detected
only in the visual band (75\% of the total) are fainter than the UV detection
limit, more reddened, or older than the others.\\  
4. Most of the clusters with measurable size have radius $< 5$ pc. When combined
with the age or mass estimates above, this implies they are bound, and they 
will evolve into objects similar to globular clusters.\\
5. UV-selected circumnuclear rings occur preferentially in barred galaxies
of Hubble type S0 to Sc. There is no clear correspondence between their 
occurrence and that of a Seyfert or LINER-type AGN.

Further observations in additional wavebands at the {\it HST} resolution
can significantly constrain the nature of these objects, and disentangle
the effects of age, reddening, and star-forming initial conditions
and history.

\acknowledgements
We are grateful to J. N. Bahcall for his contribution to the earlier
stages of this survey, to G. R. Meurer for helpful advice and comments,
and to the technical
staff at Lick Observatory for assistance during the spectroscopic
observations of NGC 1079.
This work was supported by grant GO-3519
 from the Space Telescope
Science Institute, which is operated by AURA, Inc., under NASA
contract NAS 5-26555. A. V. F., L. C. H., and A. J. B. also acknowledge
support by grant AR-5291 from the Space Telescope
Science Institute.
D. M., A. S., and H. -W. R acknowledge support by  the U.S.-Israel
Binational Science Foundation grant 94-00300.

\newpage

\begin{figure}
\epsscale{1.1}
\plotone{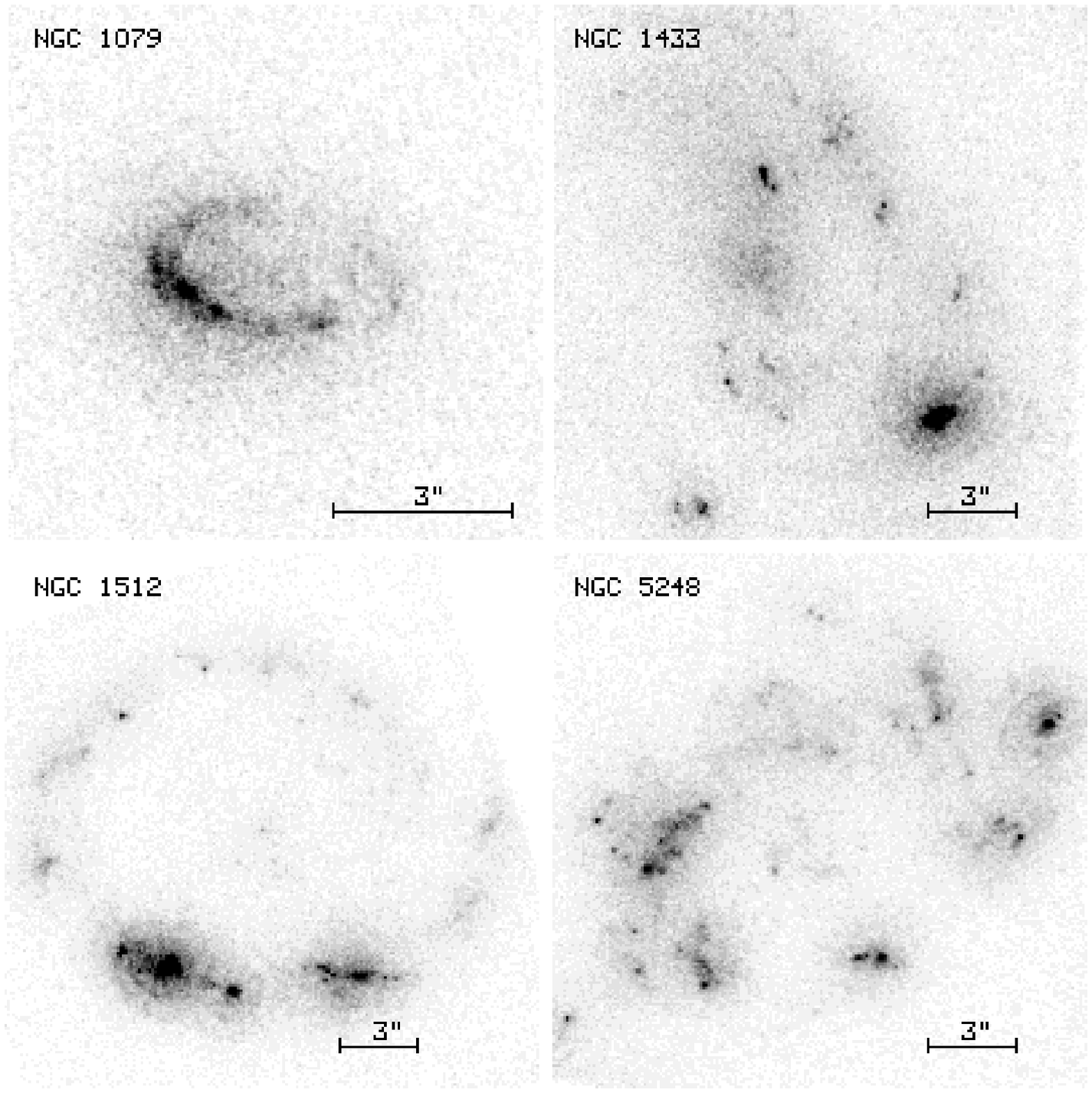}
\figcaption{{\it HST} FOC F220W images of four of the circumnuclear rings. North is
up and east is to the left. The scale of each image section is marked.}
\end{figure}

\begin{figure}
\epsscale{1.1}
\plotone{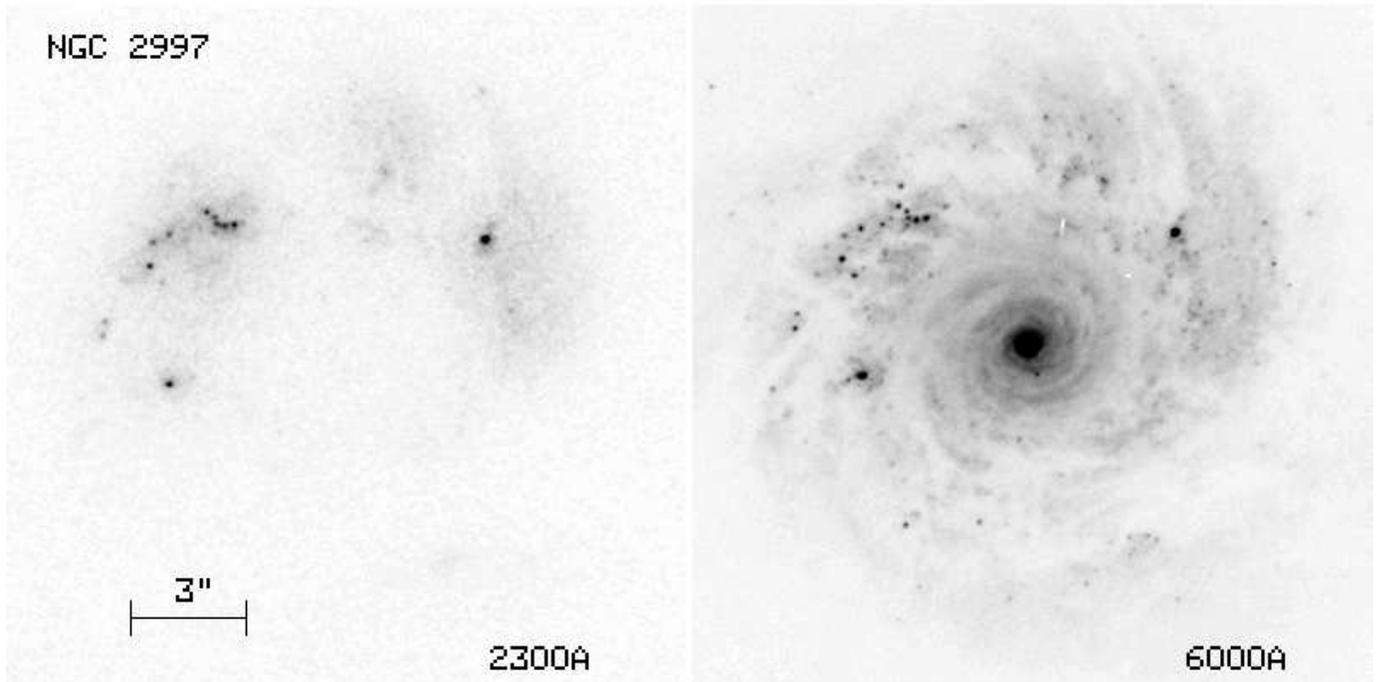}
\figcaption{{\it HST} images of the nuclear ring in NGC 2997. Left panel: FOC F220W
(UV) image. Right panel: WFPC2 F606W (optical) image, to the same scale.
 Note the different appearance
of the galaxy in the UV image, which highlights only regions of current and recent 
star formation. North is up, east is to the left.}
\end{figure}

\begin{figure}
\epsscale{1.0}
\plotone{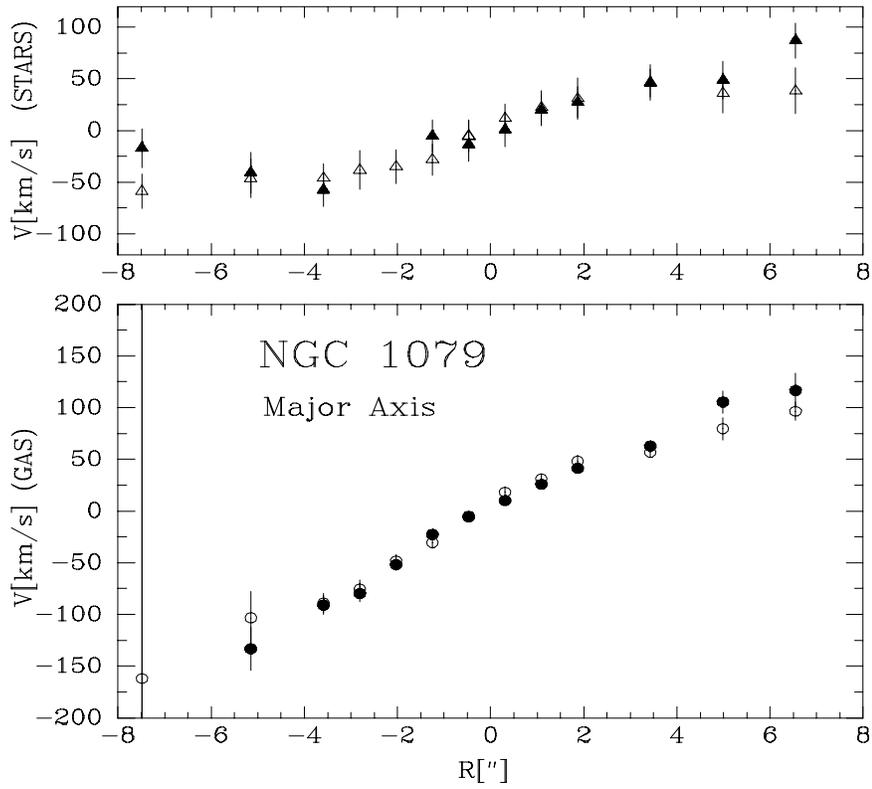}
\figcaption{Major-axis rotation curves obtained for NGC 1079 using
absorption lines, which trace stars (top panel), and emission
lines, which trace gas (bottom panel). Filled and empty symbols
denote two independent observations.}
\end{figure}

\begin{figure}
\epsscale{1.3}
\plotone{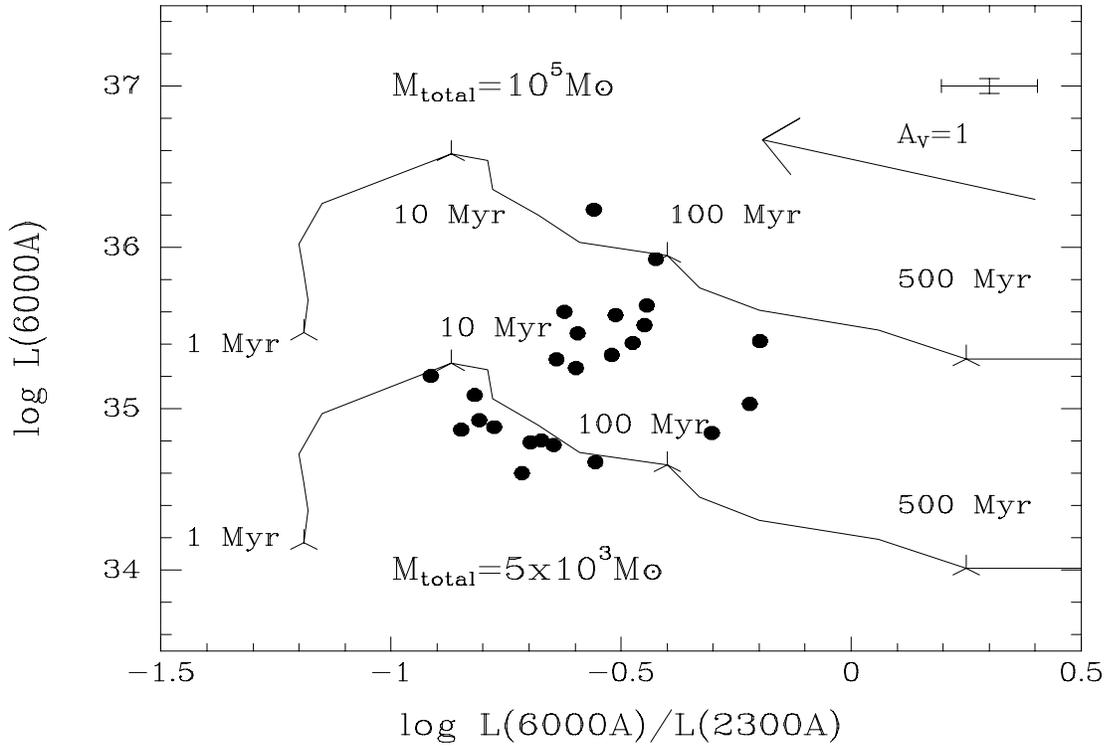}
\figcaption{Color-magnitude diagram for the clusters in NGC 2997,
relating the 6000 \AA\ luminosity to the 6000 \AA\ -- 2300 \AA\ color.
The points show the color and luminosity of all the NGC 2997 clusters
detected in both bands, as listed in Table 7.
 The two curves
show how two clusters of different total mass move on the diagram with
time.
An exponentially decreasing star formation rate with characteristic
 decay time of 5 Myr is assumed.
 Also shown (arrow) is a de-reddening vector, giving the correction
for one magnitude of foreground visual extinction, and the typical error bar
for an observed point (upper-right corner).}
\end{figure}

\begin{figure}
\epsscale{1.3}
\plotone{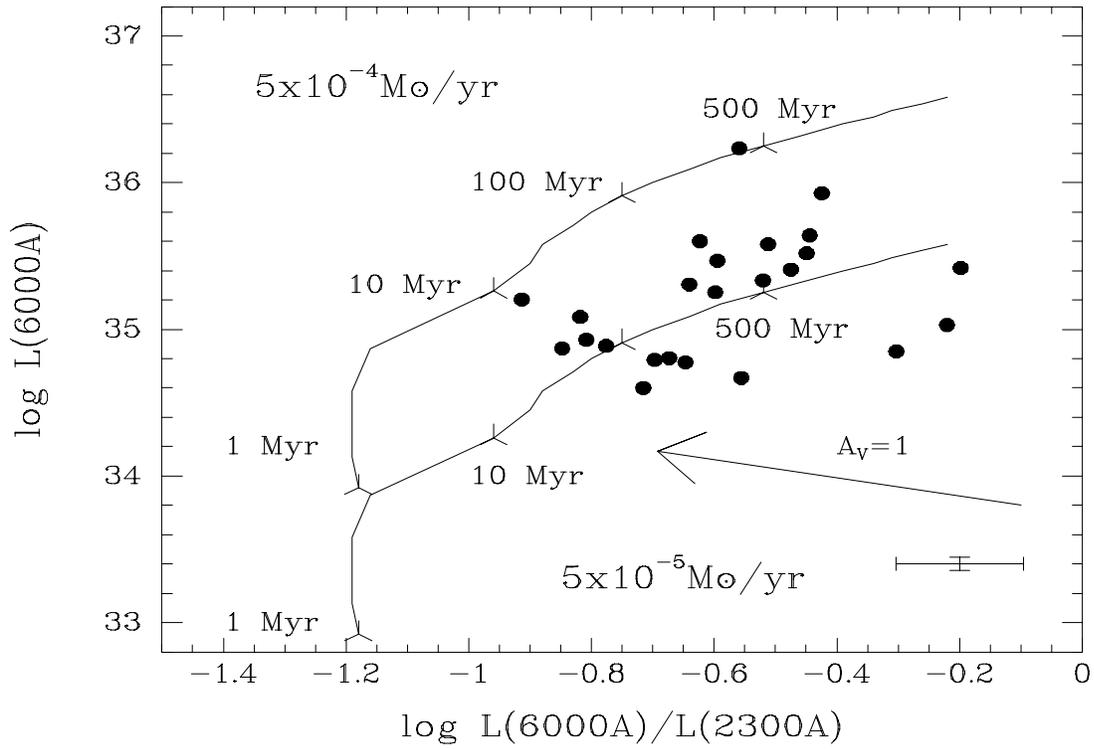}
\figcaption{As in Fig. 4, but with the modelled evolution of two clusters
with a {\it constant} ($5\times 10^{-4}$ and $5\times 10^{-5} 
M_{\odot}$ yr$^{-1}$) star formation rate (SFR). The mass of the cluster
in stars at a given time is the SFR times the age.}
\end{figure}

\begin{figure}
\epsscale{1.2}
\plotone{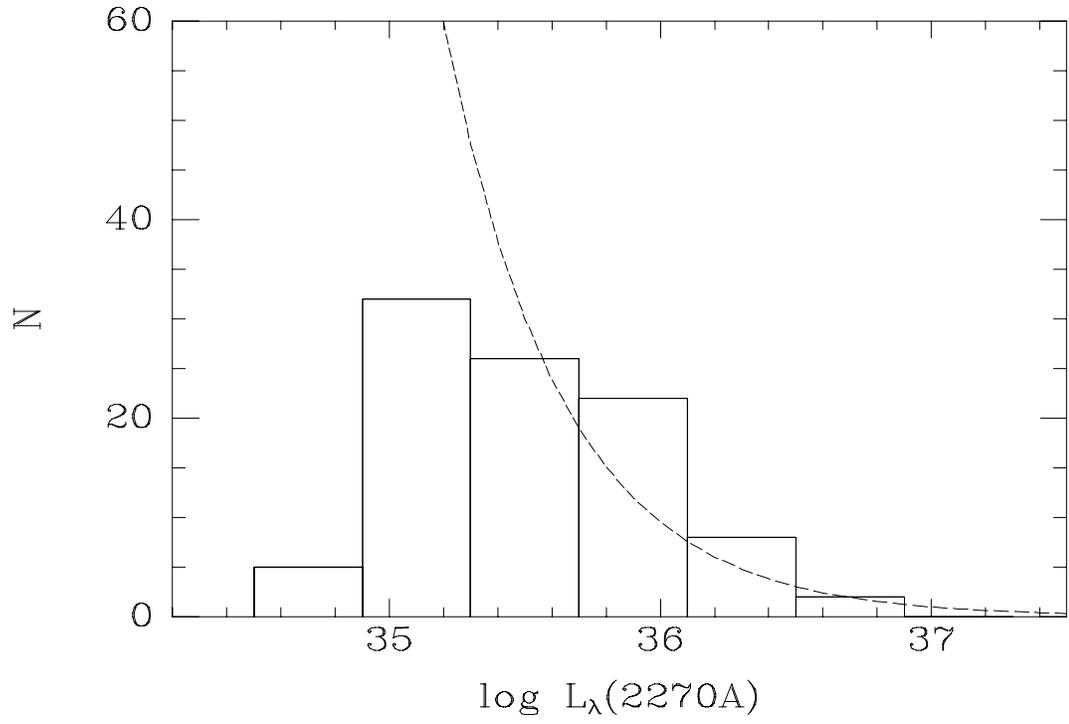}
\figcaption{Combined observed UV luminosity function for the clusters in
NGC 1433, NGC 1512, and NGC 2997. Also plotted (dashed line) is
a luminosity function of the form $N\propto L^{-2}$. }
\end{figure}

\begin{figure}
\epsscale{1.2}
\plotone{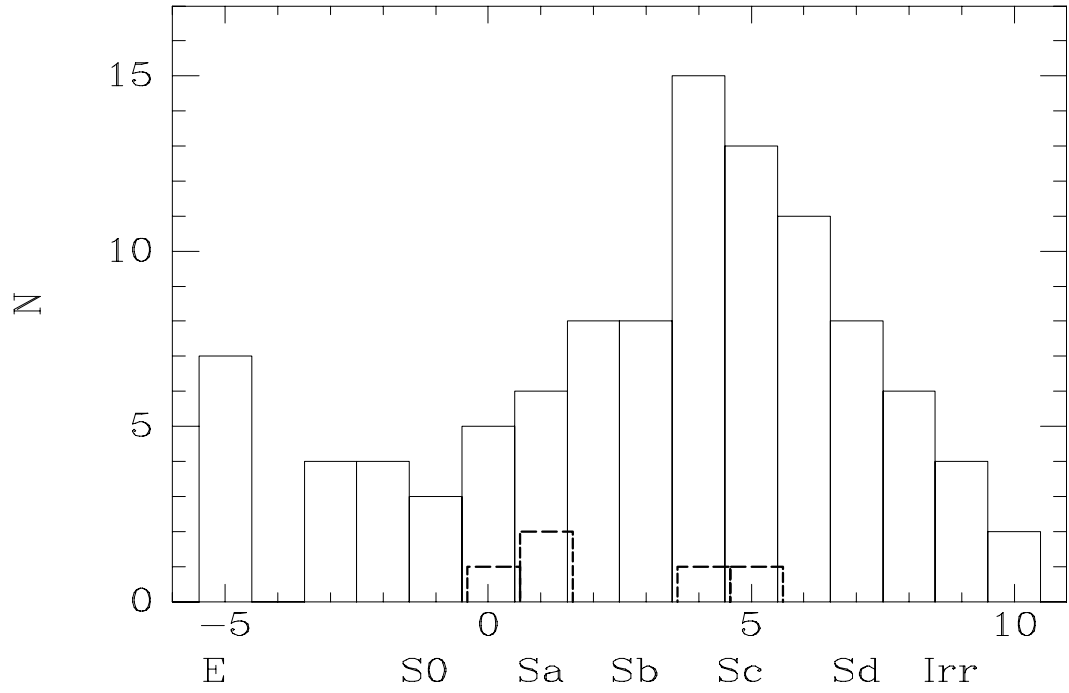}
\figcaption{The distribution of morphological types among the
103 galaxies in the {\it HST} survey, and among the five nuclear-ring
galaxies (dashed histogram), based on de Vaucouleurs's ``T-type''
classification.}
\end{figure}

\end{document}